\documentclass[prb,aps,graphics,epsfig,floats,showpacs,twocolumn,superscriptaddress]{revtex4}
\usepackage{graphicx}
\usepackage{amssymb}
\usepackage{amsmath}
\usepackage{color}

\begin{document}

\title{Angular characterization of spin-orbit torque and thermoelectric effects}

\author{Huanglin Yang$^{1}$, Huanjian Chen$^{1}$, Meng Tang$^{1}$, Shuai Hu$^{1}$ and Xuepeng Qiu}\email{xpqiu@tongji.edu.cn}

\affiliation{Shanghai Key Laboratory of Special Artificial Microstructure Materials and Technology and Pohl Institute of Solid State Physics and
School of Physics Science and Engineering, Tongji University, Shanghai 200092, China}

\date{\today}
\vspace{5cm}

\begin{abstract}
Arising from the interplay between charge, spin and orbital of electrons, spin-orbit torque (SOT) has attracted immense interest in the past decade. Despite vast progress, the existing quantification methods of SOT still have their respective restrictions on the magnetic anisotropy, the entanglement between SOT effective fields, and the artifacts from the thermal gradient and the planar Hall effect, $etc$. Thus, accurately characterizing SOT across diverse samples remains as a critical need. In this work, with the aim of removing the afore-mentioned restrictions, thus enabling the universal SOT quantification, we report the characterization of the sign and amplitude of SOT by angular measurements. We first validate the applicability of our angular characterization in a perpendicularly magnetized Pt/Co-Ni heterostructure by showing excellent agreements to the results of conventional quantification methods. Remarkably, the thermoelectric effects, \emph{i.e.}, the anomalous Nernst effect (ANE) arising from the temperature gradient can be self-consistently disentangled and quantified from the field dependence of the angular characterization. The superiority of this angular characterization has been further demonstrated in a Cu/CoTb/Cu sample with large ANE but negligible SOT, and in a Pt/Co-Ni sample with weak perpendicular magnetic anisotropy (PMA), for which the conventional quantification methods are not applicable and even yield fatal error. By providing a comprehensive and versatile way to characterize SOT and thermoelectric effects in diverse heterostructures, our results pave the important foundation for the spin-orbitronic study as well as the interdisciplinary research of thermal spintronic.
\end{abstract}
\maketitle
\section{\label{sec:level1}Introduction
	\protect
	}
\indent Spin-orbit torque (SOT) has attracted intense interests for efficient electrical manipulation of magnetization which is at the root of numerous spintronic applications\cite{Qiu-AM}. In heavy metal (HM)/ferromagnet (FM) heterostructure or ferromagnet lacking inversion symmetry, SOT emerges due to the spin-orbit related mechanisms such as the spin Hall effect\cite{Liu-Sci2012,Liu-PRL2012,Haazen-NatM2013,Sinova-RevMP2015,Kato_Science2004,Qiu-SciRep}, the Rashba effect\cite{Miron-Nat, Manchon_NatM2015} or the topological quantum effect\cite{Hasan_RMP2010,Moore_Nat2010,Qi_RMP2011,Mellnik-Nat2014}. Compared to its counterpart of conventional spin-transfer torque (STT), SOT sheds greater prospects in the efficiency and speed\cite{Ohno-NatM2012,Miron-NatM2010,Emori_NatM2013,Ryu_NatNa2013}. In STT, the efficiency is proportional to the spin polarization $P$ of ferromagnet which is essentially smaller than 1. In contrast, SOT scales with the spin Hall angle ${\theta_{SH}}$ without upper limit which can be as large as hundreds in topological insulator\cite{FanY_NatM2014,Dc_NatM2018,Mellnik-Nat2014}. Moreover, the prior study has demonstrated the SOT drives magnetization switching in the hundred pico-second timescale that is much faster than that in the state-of-art STT device\cite{Garello_APL2014}. Additionally, in a magnetic tunneling junction, the writing current of SOT only passes through the underlayer, thus avoiding the electrical breakdown of the oxide tunneling barrier\cite{Seo_IEEE2018, Oboril_IEEE2015}. With advantages in all these aspects, SOT has brought revolutionary opportunities for spin memory and logic applications.\\
\indent Tracing the history of development, the recognition of SOT has been continuously revised for both the direction and magnitude. At the early phase of the study, SOT effective field in HM/FM was considered as a quasi-Rashba field that transverse to the current and in the film plane\cite{Pi-APL2010,Miron-NatM2010,Miron_NatM2011,Suzuki_APL2011}, but later on, it was recognized to be an out-of-plane effective field as inferring from the first perpendicular magnetization switching experiment\cite{Slonczewski_PRB2010}. On the other hand, the magnitude has also been revised from 1 to $10^{-2}~Tesla$ at the current density of 10$^8~A/cm^2$ \cite{Miron_NatM2011,Pi-APL2010}. At present, it is understood that SOT takes complex form which should be discomposed into two orthogonal components \cite{Kim-NatM2012,Garello_NatN2013,Haney_PRB2013,Khvalkovskiy_PRB2013,Qiu-SciRep,Avci-PRB2014,Qiu-Natno2015,Hayashi_PRB2014_Quantitativea}: a longitudinal effective field $\overrightarrow{H_L}  \parallel \overrightarrow{m} \times \overrightarrow{y}$ and a transverse effective field $\overrightarrow{H_T}  \parallel \overrightarrow{y}$, where $\overrightarrow m $ is the magnetization unit vector and $\overrightarrow y $ is the in-plane axis transverse to the current flow $\overrightarrow x $ direction. These evolving recognitions for SOT have been radically moved forward by the advances in SOT characterization.\\
\indent Although a variety of electrical transport methods for characterizing SOT effective fields have been developed\cite{Pi-APL2010,Kim-NatM2012,Garello_NatN2013,Hayashi_PRB2014_Quantitativea,Qiu-SciRep,Avci-PRB2014,Liu-PRL2011}, most of them have their respective restrictions that hindering the comprehensive and accurate SOT characterization. The first restriction is due to the complex form and entanglement of SOT effective fields. Both the SOT effective fields $\overrightarrow{H_{L}}$ and $\overrightarrow{H_{T}}$ could contribute to the electrical signal, thus requiring careful disentanglement and quantification. The SOT characterization methods, such as the DC\cite{Slonczewski_PRB2010,Liu-PRL2012,FanXin_NC2013} and ST-FMR ones\cite{Liu-PRL2011,Mellnik-Nat2014,WangY-PRL2015}, mostly only consider one specific SOT component by neglecting another SOT component. The second restriction is for the magnetic anisotropy. In the perpendicular magnetized HM/FM heterostructures which are of the main interests of SOT research, the models of SOT characterization are mostly built on the conditions of coherent magnetization rotation and saturated magnetic state, and therefore a strong perpendicular magnetic anisotropy is required. Lastly, the thermoelectric effect has been largely neglected in the existing SOT characterization methods\cite{Pi-APL2010,Pai_PRB2016,YangYM_PRB2016,Avci-PRB2014,Liu-PRL2012}. As we will demonstrate below, the contribution from the ANE can be very large and even dominates the electric signal. It should be noted that not only to quantify SOT accurately, the appropriate characterization and utilization of this thermoelectric effect can also bring great opportunity for expanding the horizon of spintronic research and application. Almost at the same time as the emergence of the SOT study, Slonczewski has proposed the thermal initiation of spin-transfer torque from magnons\cite{Slonczewski_PRB2010}. Later on, several groups have further demonstrated the thermal creation of spin current and spin torque in magnetic heterostructures through the magnons, spin-dependent Seebeck and spin-Nernst effect (SNE)\cite{XiaKe_PRL2011,Bauer_NatM2012,Yu_PLA2016}.\\
\indent In this work, a new angular characterization method with least restrictions to study the SOT and thermoelectric effects in various magnetic heterostructures. By rotating the sample under a strong magnetic field, the simultaneously recorded harmonic Hall voltages are utilized to analytically derive the SOT effective fields and the ANE contribution across the sample. After giving the principal equations that in seccession of previous work by Avci \emph{et al.}\cite{Avci-PRB2014}, we first analyze and simulate the angular dependences of harmonic signal originated from SOT and ANE. Subsequently, the applicability of the angular characterization is verified by comparing the results with other established quantification methods in a perpendicularly magnetized Pt/Co-Ni sample. At last, we apply the angular characterization for two special samples, for which the other conventional quantification methods are not applicable due to the restrictions of large thermoelectric effect and weak PMA. By accurately characterizing SOT and thermoelectric effects with high sensitivity, simple procedures and wide applicability, our works not only provide an important basis for SOT research but also allow the further investigation of the interplay between SOT and thermoelectric effects.\\
\section{Model of Angular Characterization: SOT and Thermoelectric Effects
\protect
}
\indent We first describe the model of angular characterization and the corresponding equations in detail. In a NM/FM heterostructure, the measurements are performed by applying an AC current $I_{AC} = {I_0}\sin (\omega t)$ along the $\overrightarrow{x}$ direction. The first and second harmonic Hall voltages are measured simultaneously along the $\overrightarrow{y}$ direction by using two lock-in amplifiers. Due to the AC current-induced periodic SOT effective fields, the magnetization oscillates around its equilibrium state which in turn generates the harmonic Hall voltages. Additionally, the thermoelectric effect also contributes a second harmonic Hall voltage through ANE.\\
\subsection{SOT Effect
\protect
}
\begin{figure}
	\includegraphics[width=8cm]{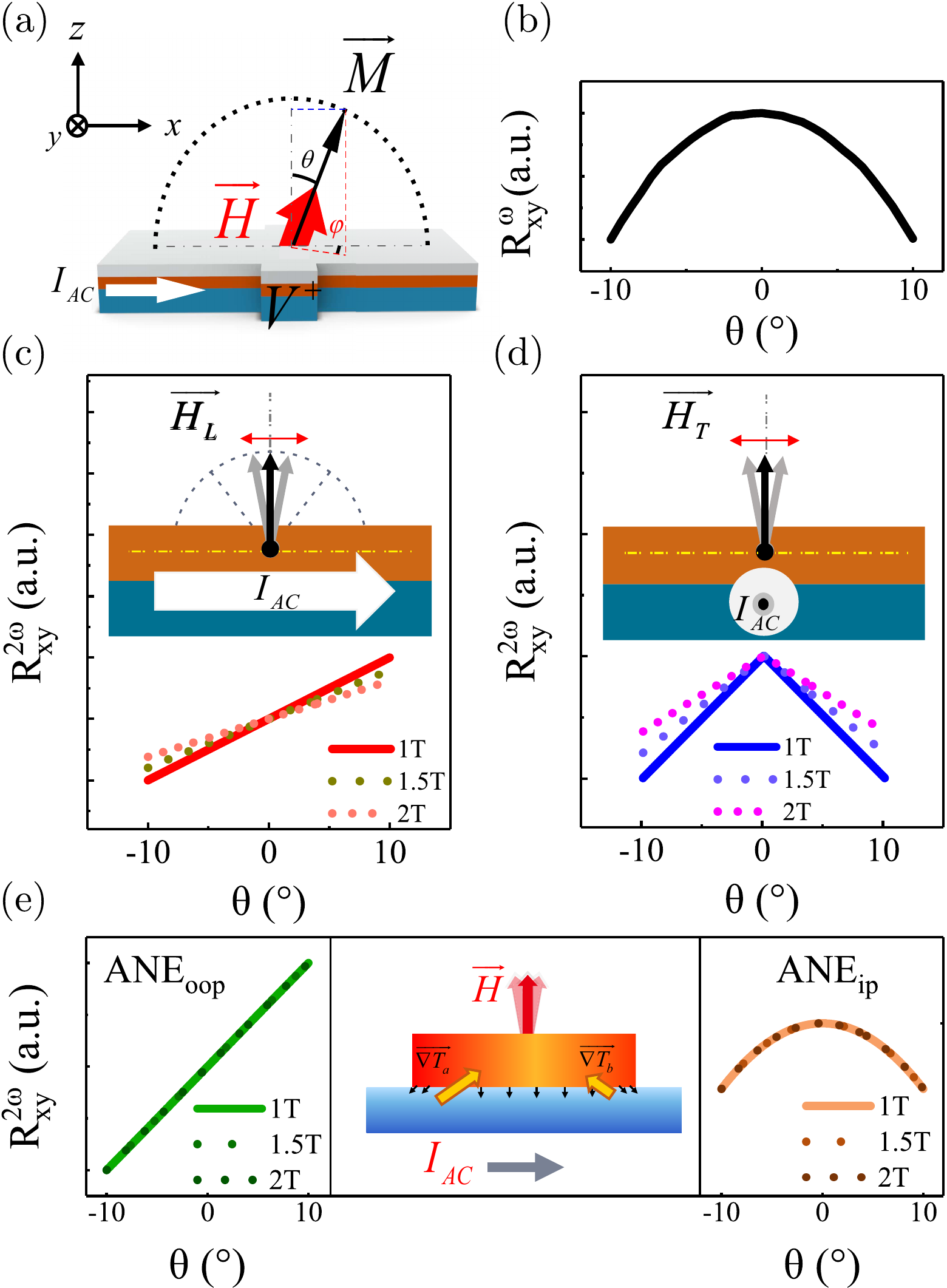}
	\caption{\label{fig1} (Color online) Simulation for longitudinal scheme. (a) The angular measurement in the $xz$ plane around the $z$-axis. (b) The first harmonic Hall resistance characterizes the magnetization equilibrium position. The second harmonic Hall resistance characterizes the magnetization oscillation due to (c) a longitudinal effective field (${H _{L}}$), and (d) a transverse effective field (${H _{T}}$) with different external magnetic field. (e) Schematic of the out of plane and in-plane thermal gradient caused by Joule heating (center) and angular dependences of the second harmonic signal from the out of plane and in-plane ANE with different external magnetic field (left and right).}
\end{figure}
\begin{figure}
	\includegraphics[width=8cm]{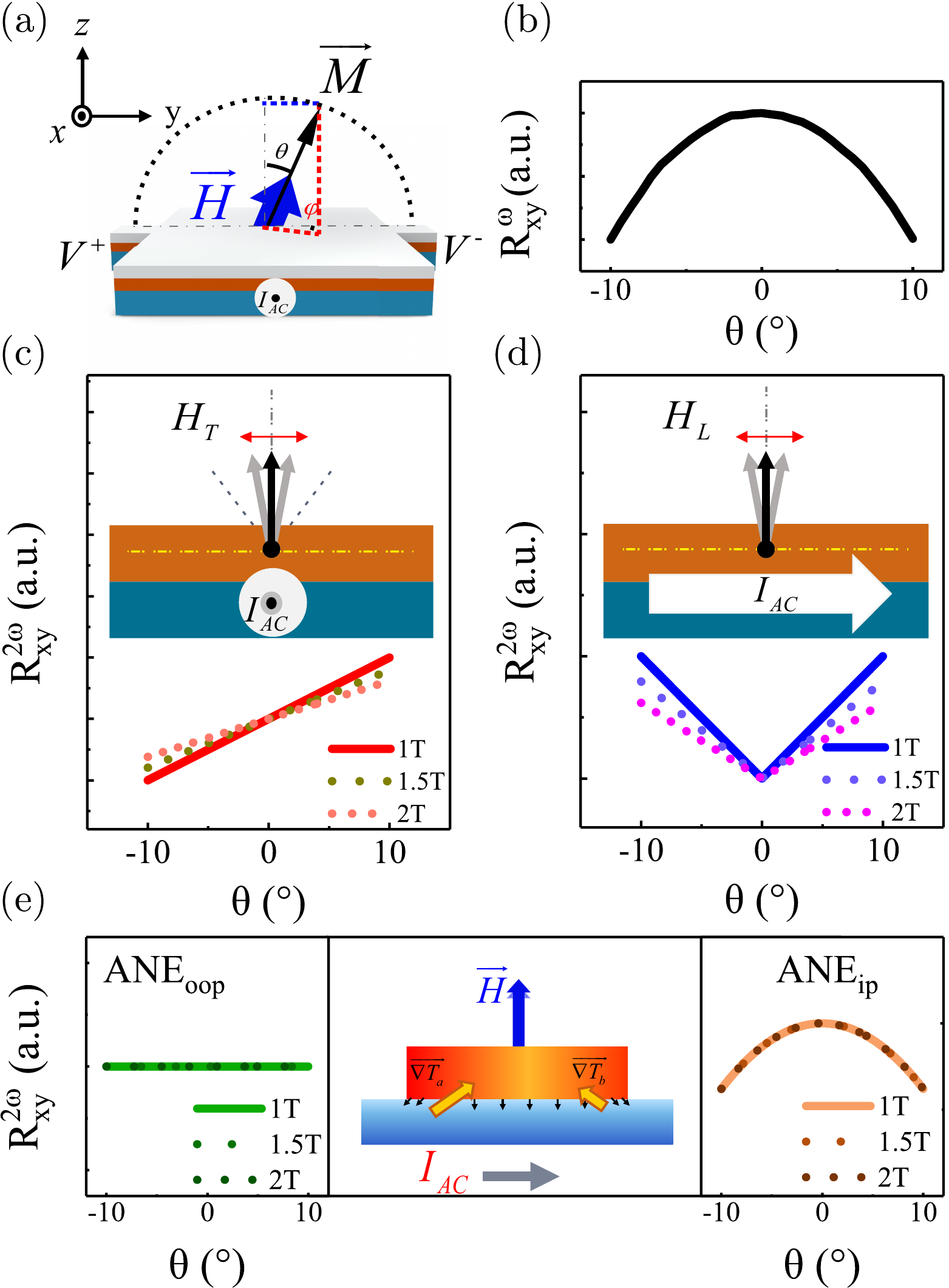}
	\caption{\label{fig2}(Color online) Simulation for transverse scheme. (a) The angular measurement in the $yz$ plane around the $z$-axis. (b) The first harmonic Hall resistance characterizes the magnetization equilibrium position. The second harmonic Hall resistance characterizes the magnetization oscillation due to (c) a transverse effective field (${H _{T}}$), and (d) a longitudinal effective field (${H _{L}}$) with different external magnetic field. (e) Schematic of the out of plane and in-plane thermal gradient caused by Joule heating (center) and angular dependences of the second harmonic signal from the out of plane and in-plane ANE with different external magnetic field (left and right).}
\end{figure}
\indent For SOT measurements, the harmonic Hall voltages are measured in both the longitudinal and transverse schemes.
In the longitudinal scheme, the magnetization of the sample is rotated in the $xz$ plane with a constant external magnetic field $\overrightarrow {H_{ext}}$, as shown in Fig.~\ref{fig1}(a, b). The magnetization oscillation is quantified into $\Delta \theta$ and $\Delta \varphi $, which are caused by the $\overrightarrow {H_{L}} $ and the $\overrightarrow {H_{T}} $ respectively, as shown in Fig.~\ref{fig1}(c, d).
In the transverse scheme, the magnetization of the sample is rotated in the $yz$ plane with a constant external magnetic field $\overrightarrow {H_{ext}}$, as shown in Fig.~\ref{fig2}(a, b). In contrast to the longitudinal scheme, the $\overrightarrow {H_{T}} $ causes the $\Delta \theta$ and the $\overrightarrow {H_{L}} $ causes the $\Delta \varphi $ which can be seen from Fig.~\ref{fig2}(c, d).
By considering both the AHE and PHE, the Hall voltage can be written as
\begin{equation}\label{eq:Hall voltage}
	 \begin{split}
&{V_{xy}}(t) = {I_0}\sin \omega t{R_{AHE}}\cos (\theta+\Delta \theta \sin \omega t)\\
&+ {I_0}\sin \omega t{R_{PHE}}{\sin ^2}(\theta  + \Delta \theta \sin \omega t)\\
&\sin (2(\varphi  + \Delta \varphi \sin \omega t))\\
	\end{split}
\end{equation}
where $\theta $ and $\varphi $ define the magnetization equilibrium direction.\\
\indent As $\theta $ and $\varphi $ are determined by $\overrightarrow {{H_k}} $ (anisotropic field) and $\overrightarrow {{H_{ext}}} $, $\Delta \theta $ and $\Delta \varphi $ represented magnetization oscillations are determined by $\overrightarrow {H_{{I_0}}}={\overrightarrow {H_{{L}}}}+\overrightarrow {H_{{T}}}$ (current-induced effective field), the first Harmonic Hall resistances can be expressed as:
\begin{equation}\label{eq:1st Hall resistance}
\begin{split}
R_{xy}^{\omega } = {R_{AHE}}\cos (\theta ) + {R_{PHE}}{\sin ^2}(\theta )\sin (2\varphi )
\end{split}
\end{equation}
And the second Harmonic Hall resistances can be expressed as:\\
\indent For the longitudinal scheme,
 \begin{equation}\label{eq:N2}
	\begin{split}
		R_{xy}^{2\omega } &= { \frac{1}{2}{R_{AHE}}\frac{H_{L}}{{H_{ext}}+{H_k}}}\sin \theta\\
		&-{{R_{PHE}}\frac{H_{T}}{{H_{ext}}+{H_k}}}\sin^2\theta\\
	\end{split}
\end{equation}
\indent For the transverse scheme,
\begin{equation}\label{eq:N22}
	\begin{split}
		R_{xy}^{2\omega } &= { \frac{1}{2}{R_{AHE}}\frac{H_{T}}{{H_{ext}}+{H_k}}}\sin \theta\\
		&+{{R_{PHE}}\frac{H_{L}}{{H_{ext}}+{H_k}}}\sin^2\theta\\
	\end{split}
\end{equation}
\subsection{Thermoelectric Effect
\protect
}
\begin{figure}
	\includegraphics[width=8cm]{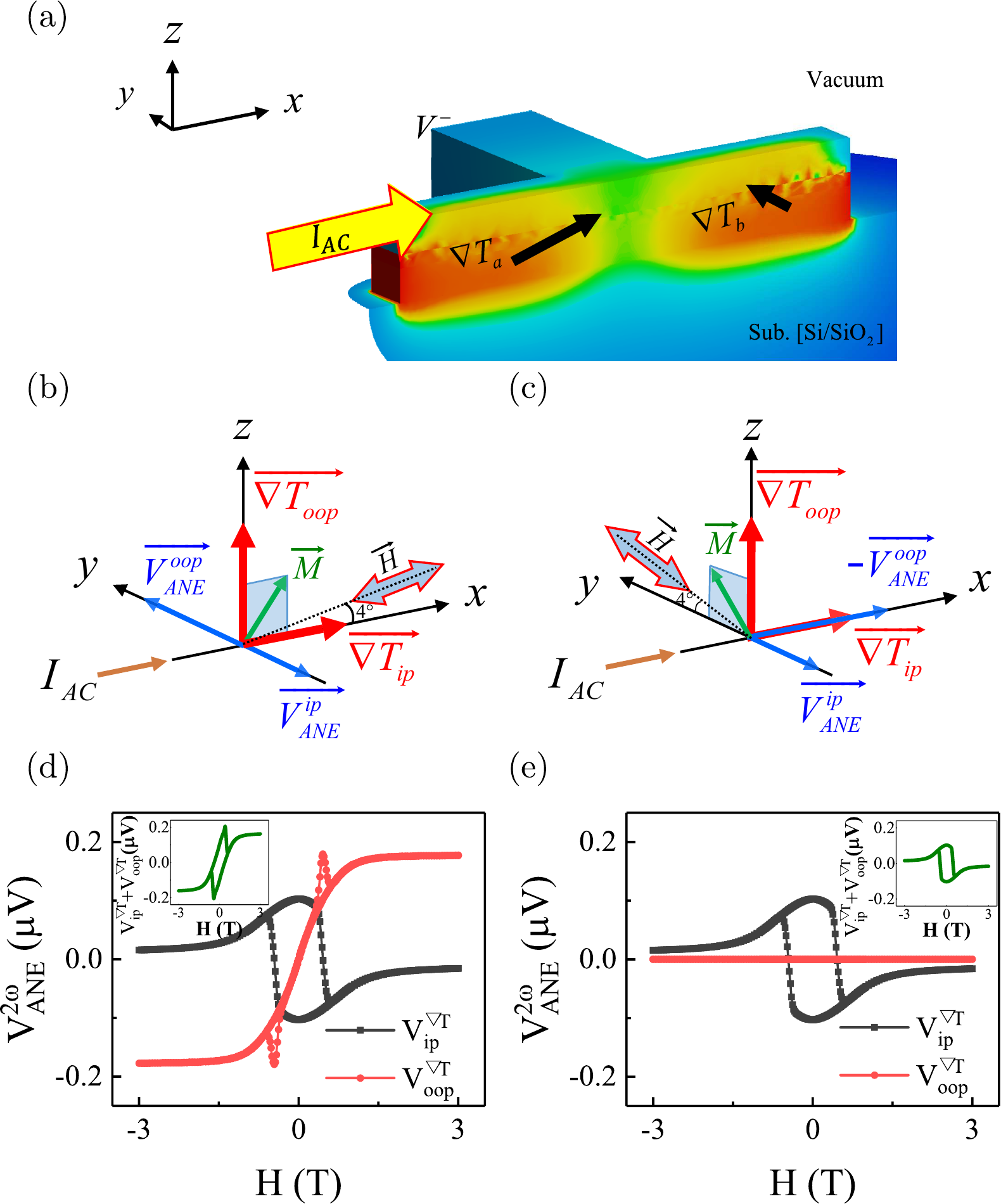}
	\caption{\label{fig3} (Color online) (a) The finite element analysis of thermal gradient caused by the current. (b) The analysis of ANE signals for the longitudinal (b) and transverse (c) schemes. The simulation of in-plane and out of plane ANE signals with scanning external field at the full $H$ range\cite{Qiu-SciRep} in the longitudinal (d) and transverse (e) schemes. The insets in (d, e) show the total ANE loops.
}
\end{figure}
\indent Additionally, the measured second harmonic signal also contains the thermoelectric contribution. Fig.~\ref{fig3} (a) shows the current induced thermal gradient ($\nabla T_a, \nabla T_b$) in the Hall bar device by using the finite element analysis [details see Appendix]. The thermal gradient is correlated with the quadratic square of the current. Therefore, for an AC current, one has
	\begin{equation}\label{eq:thermal Gradient}
	\begin{split}
	{\nabla T \propto {I^2}{R_s} = I{{}_0^2}{\sin ^2}(\omega t){R_s}}.
	\end{split}
	\end{equation}
\begin{equation}\label{eq:thermal ANE}
\begin{split}
\overrightarrow{V_{ANE}} \propto  - \alpha \overrightarrow {\nabla T}  \times \overrightarrow m \propto{\sin ^2}(\omega t){R_s}
\end{split}
\end{equation}
where $R_s$ is the sample resistance. And we can clearly see that the ANE is manifested as a second harmonic term.
Furthermore, the thermal gradient can be decomposed into the in-plane and out of plane parts in the Cartesian coordinates, as shown in Fig.~\ref{fig3}(b, c). The vector of the thermal gradient are defined as ${\overrightarrow {\nabla T} _{ip}} = \nabla {T_{ip}}(1,0,0)$ (the in-plane part), ${\overrightarrow {\nabla T} _{oop}} = \nabla {T_{oop}}(0,0,1)$ (the out of plane part), respectively. Fig.~\ref{fig3}(b, c) shows the analysis of ANE in the longitudinal and transverse schemes. When a magnetization $\overrightarrow m$ experiences a thermal gradient ${\overrightarrow {\nabla T}}$, the ANE voltage is produced along the direction of $\overrightarrow {{V_{ANE}}} \parallel \overrightarrow {\nabla T}  \times \overrightarrow m $. Thus, in the longitudinal scheme, the in-plane and out of plane ANE voltages are both along the $\overrightarrow y$ direction. In the transverse scheme, the direction of in-plane ANE voltage lies in $\overrightarrow y$ while the out of plane one lies in $\overrightarrow x$.\\
\indent As described in Appendix, the second harmonic ANE contribution can be written as
\begin{equation}\label{eq:ANE_oop}
\begin{split}
R_{oop,ANE}^{2\omega } = {I_0}\alpha \nabla {T_{oop}}\sin \theta \cos \varphi
\end{split}
\end{equation}
and
\begin{equation}\label{eq:ANE_ip}
\begin{split}
R_{ip,ANE}^{2\omega } = - {I_0}\alpha \nabla {T_{ip}}\cos \theta
\end{split}
\end{equation}

\indent Therefore, together with the ANE analysis in Fig.~\ref{fig3}(b, c), the ANE contributions can be determined in both the two schemes. For the longitudinal scheme,
\begin{equation}\label{eq:ANE_Long}
\begin{split}
R_{\nabla T}^{2\omega } = {I_0}\alpha (\nabla {T_{oop}}\sin \theta  - \nabla {T_{ip}}\cos \theta )
\end{split}
\end{equation}
And for the transverse scheme,
\begin{equation}\label{eq:ANE_Trans}
\begin{split}
R_{\nabla T}^{2\omega } =  - {I_0}\alpha \nabla {T_{ip}}\cos \theta
\end{split}
\end{equation}
\indent Fig.~\ref{fig3}(d, e) show the simulated hysteresis loops of ANE voltage for a perpendicular magnetized Si/SiO$_2$/Pt (4)/Co (0.7)/Pt (1)/SiO$_2$ (3) device, in which the magnetization direction is obtained from the first harmonic Hall voltage by canning external field at full $H$ range with a tilt angle $\theta _H = 86 ^\circ$\cite{Qiu-SciRep}.
For small H range, the more specific ANE simulation curves are shown in Fig.~\ref{fig1}(e) and Fig.~\ref{fig2}(e). One can find that the ANE contributions remain unchanged with the external magnetic field strength.\\
\indent Combing the SOT effect and thermoelectric contribution, the second harmonic Hall resistance can be expressed as:\\
\indent For the longitudinal scheme,
 \begin{equation}\label{eq:Q_Long}
\begin{split}
R_{xy}^{2\omega } &= { \frac{1}{2}{R_{AHE}}\frac{H_{L}}{{H_{ext}}+{H_k}}}\sin \theta\\
&-{{R_{PHE}}\frac{H_{T}}{{H_{ext}}+{H_k}}}\sin^2\theta\\
& + {I_0}\alpha (\nabla {T_{oop}}\sin \theta-\nabla {T_{ip}}\cos \theta)
\end{split}
\end{equation}
\indent For the transverse scheme,
\begin{equation}\label{eq:Q_Trans}
\begin{split}
R_{xy}^{2\omega } &= { \frac{1}{2}{R_{AHE}}\frac{H_{T}}{{H_{ext}}+{H_k}}}\sin \theta\\
&+{{R_{PHE}}\frac{H_{L}}{{H_{ext}}+{H_k}}}\sin^2\theta\\
& - {I_0}\alpha \nabla {T_{ip}}\cos \theta
\end{split}
\end{equation}
\indent Moreover, the PHE term appears to be small in general due to the small values of both PHE and ${\sin^2\theta}$ so it can be neglected. For angular characterization at small $\theta$ range, $\cos \theta$ is close to 1, thus the in-plane ANE contribution remains as a constant ($C$). Therefore, Eq.~(\ref{eq:Q_Long}) and~(\ref{eq:Q_Trans}) can be further simplified as:\\
\indent For the longitudinal scheme,
\begin{equation}\label{eq:Q_Long_simp}
\begin{split}
R_{xy}^{2\omega } &= { \frac{1}{2}{R_{AHE}}\frac{H_{L}}{{H_{ext}}+{H_k}}}\sin \theta\\
& + {I_0}\alpha \nabla {T_{oop}}\sin \theta-C
\end{split}
\end{equation}
\indent For the transverse scheme,
\begin{equation}\label{eq:Q_Trans_simp}
\begin{split}
R_{xy}^{2\omega } &= { \frac{1}{2}{R_{AHE}}\frac{H_{T}}{{H_{ext}}+{H_k}}}\sin \theta\\
& -C
\end{split}
\end{equation}
\indent For the cases that the PHE term is large and to be considered, the SOT effective fields can be corrected by the following equations:
\begin{equation}\label{eq:PHE_Correct}
\begin{split}
{H_L^{'} = H_L + 2\xi{H_T}\sin \theta }\\
{H_T^{'} = H_T - 2\xi{H_L}\sin \theta }
\end{split}
\end{equation}
where $\xi=\frac{R_{PHE}}{R_AHE}$.\\
\indent The above equations lay the important basis for quantifying SOT in this work. In the following sections, we will utilize these equations to quantify SOT and thermoelectric effect in various heterostructures.
\section{Experiment: General case and special cases
\protect
}
\subsection{General Angular Characterization of SOT in the Pt/[Co-Ni] Sample with PMA
\protect
}
\begin{figure}[tb]
\centering
	\includegraphics[width=8cm]{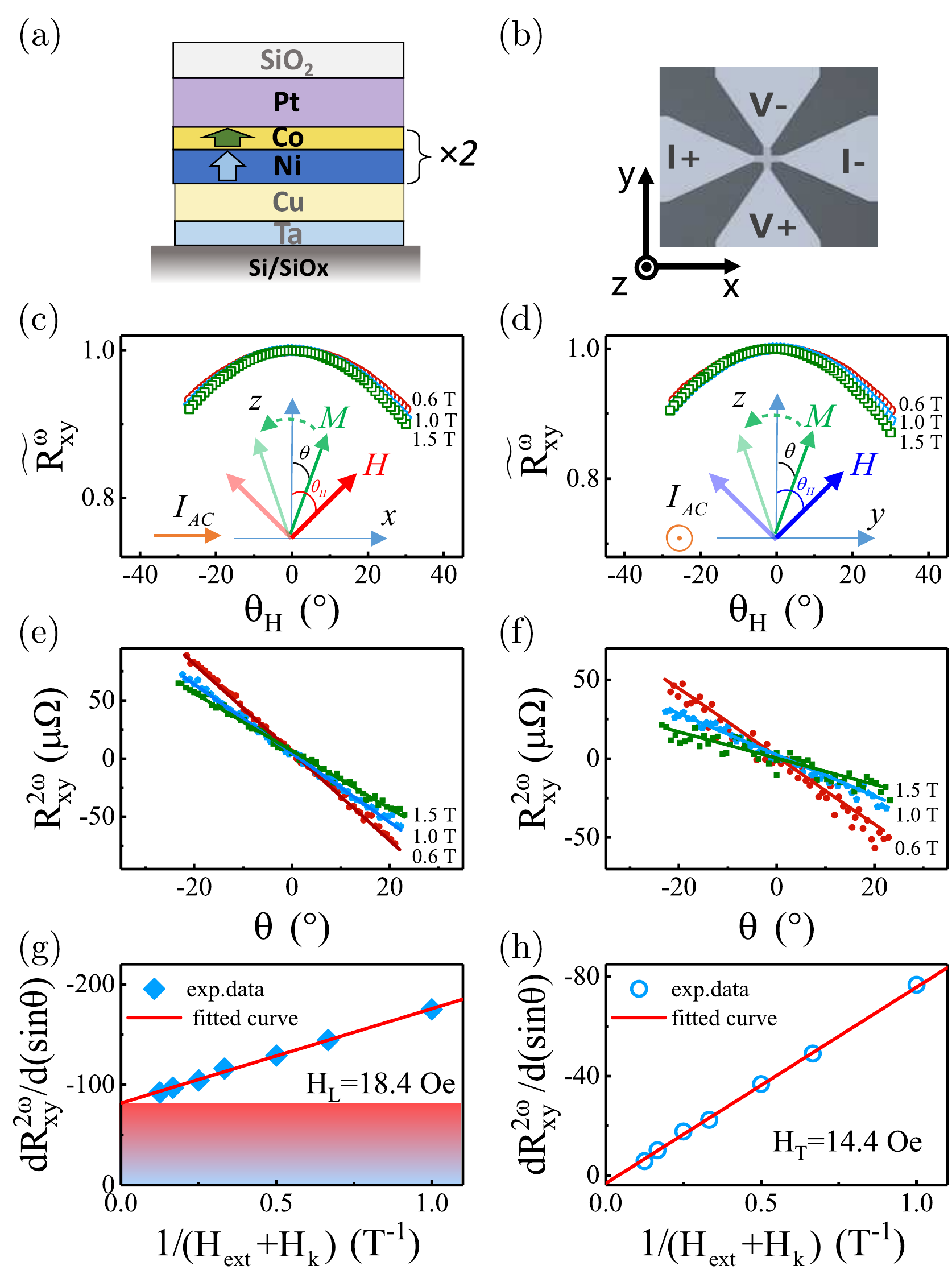}
	\caption{\label{fig4} (Color online)(a)(b) Sample structure and device microscopy image. Angular characterization of the first (c)(d) and the second (e)(f) harmonic Hall resistances with different external magnetic field (${H_{ext}} \sim 0.6~T,~1~T,~1.5~T$). (g)(h) SOT and thermoelectric contributions as a function of the inverse field ($1/({H_{ext}} + {H_k}$)). The solid lines are the linear fit of experimental data. The red rectangle box depicts the thermoelectric contributions in the second harmonic signal. In the figure, (c)(e)(g) are for the longitudinal scheme and (d)(f)(h) are for the transverse scheme.}
\end{figure}
\begin{figure}
	\includegraphics[width=8cm]{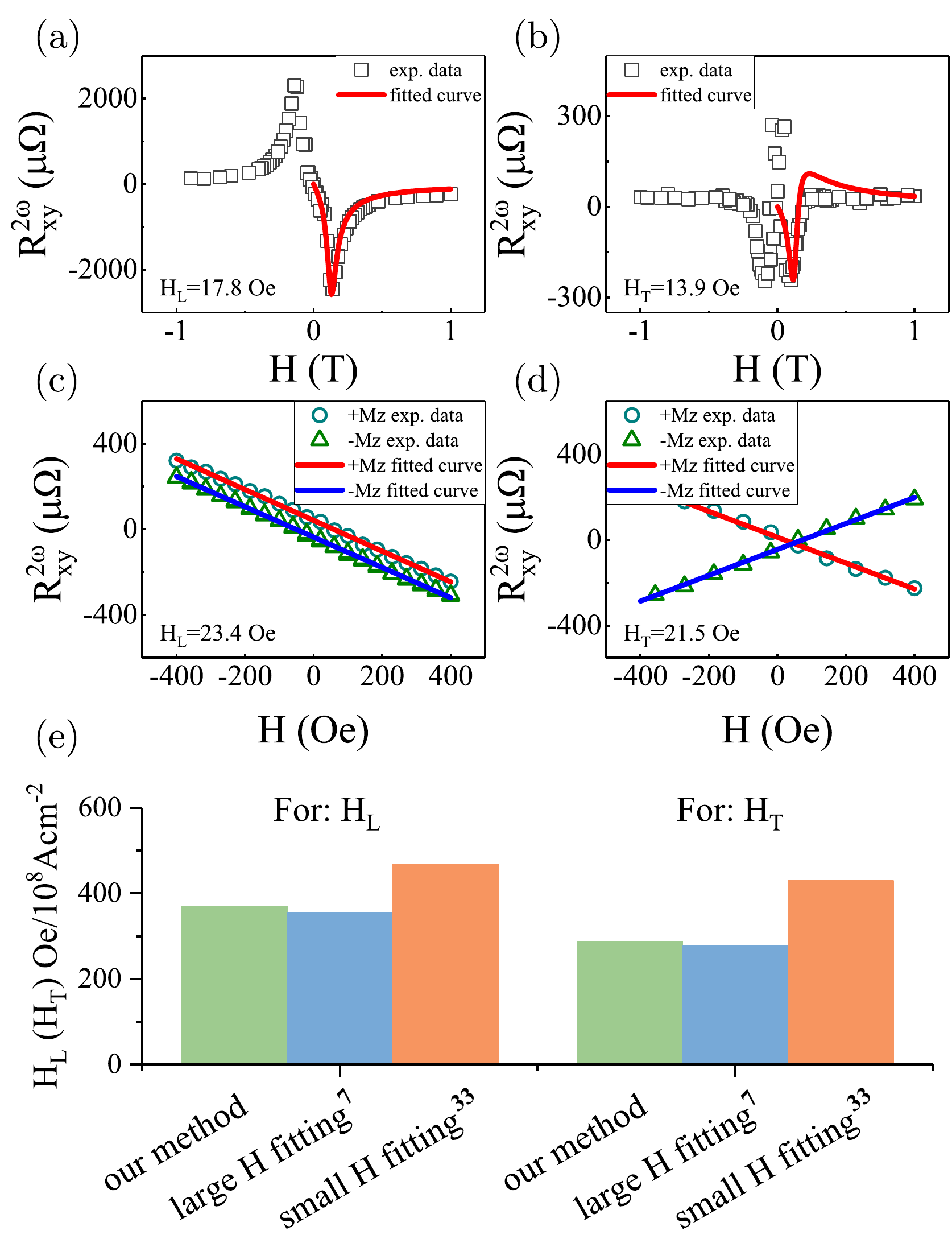}
	\caption{\label{fig5}(Color online) For the Ta/Cu/Ni-Co sample, AC harmonic characterization of SOT by fitting the second harmonic signal at the large $H$ range in the longitudinal scheme (a) and transverse scheme (b). AC harmonic characterization of SOT by fitting the second harmonic signal at the small $H$ range in the longitudinal scheme (c) and transverse scheme (d). (e) Comparison of SOT effective fields from three methods.
}
\end{figure}
\indent As a demonstration for the general case, the angular characterization is performed for a Hall bar device of Ta (2)/Cu (3)/[Co (0.3)/Ni (0.6)]$_2$/Pt (4)/SiO$_2$ (3) with PMA (number in nm). The sample and device structures are illustrated in Fig.~\ref{fig4}(a, b). All transport measurements are carried out at room temperature using Keithley 6221 current source and two lock-in amplifiers. The AC current ($f=13.7~Hz$) is applied in the $\overrightarrow x$ direction, and the first and second harmonic Hall voltages ($V_{xy}^{\omega }$ and $V_{xy}^{2\omega }$) are simultaneously recorded.\\
\indent In the measurements, the sample is rotated at $\theta_H  \in [ - 20^\circ ,20^\circ ]$ with different external magnetic field $H_{ext}$, where $\theta_H$ is the angle between $\overrightarrow{z}$ and $\overrightarrow {H_{ext}}$. Fig.~\ref{fig4}(c, d) show the angular dependence of $\widetilde {R_{xy}^\omega }$ for the longitudinal and transverse schemes, where $\widetilde {R_{xy}^\omega }$ is the normalized first harmonic Hall resistance with respect to the saturated AHE resistance. Subsequently, one can derive $\theta$ of magnetization through the conversion equation of $\theta={\rm{acos}}(\widetilde {R_{xy}^\omega })$. These measurements and procedures allow the conversion from $\theta_H$ to $\theta$, thus one can obtain the $\theta$ dependence of the second harmonic Hall resistance, as shown in Fig.~\ref{fig4}(e, f).\\
\indent We now turn to consider the derivation forms of Eq.~(\ref{eq:Q_Long_simp}) and~(\ref{eq:Q_Trans_simp}):\\
\indent For the longitudinal scheme, 
 \begin{equation}\label{eq:dVdtheta}
     \begin{split}
dR_{xy}^{2\omega }/d(\sin\theta)=\frac{{{1}}}{2}{R_{AHE}}\frac{{H_L}}{H_{ext}+H_k} + {I_0}\alpha \nabla {T_{oop}}
    \end{split}
\end{equation}
\indent For the transverse scheme, 
 \begin{equation}\label{eq:dVdtheta2}
\begin{split}
	dR_{xy}^{2\omega }/d(\sin\theta)=\frac{{{1}}}{2}{R_{AHE}}\frac{{H_T}}{H_{ext}+H_k}
\end{split}
\end{equation}
For a small $\theta$, one has $\sin\theta\approx\theta$. Therefore, $R_{xy}^{2\omega }$ can be linearly fitted with $\theta$ as represented by the straight fitting lines in Fig.~\ref{fig4}(e, f). More importantly, according to Eq.~(\ref{eq:dVdtheta}) and~(\ref{eq:dVdtheta2}), the field dependences of $dR_{xy}^{2\omega }/d(\sin\theta)$ in Fig.~\ref{fig4}(g, h) adequately quantifies both the current-induced SOT and thermoelectric effects. Note that the magnetization oscillation induced by SOT is largely suppressed at large external magnetic field, where the thermal gradient contribution is not affected by $H_{ext}$. As the PHE term is normally negligible for many SOT samples, one can characterize the $H_{L}(H_{T})$ by measuring the external field $H_{ext}$ dependence of $R_{xy}^{2\omega }$ in the longitudinal (transverse) schemes, respectively. From the $dR_{xy}^{2\omega }/d(\sin\theta) $ versus $1/({H_{ext}+H_{k}}$) in Fig.~\ref{fig4}(g, h), values of ${H_L}=18.5~Oe$, ${H_T}=14.4~Oe$, $V_{oop,ANE}^{2\omega } = 0.34~\mu V$, and $V_{ip,ANE}^{2\omega } = 0.16~\mu V$ are fitted. With further normalization, ${H_L}=370~Oe$ and ${H_T}=288~Oe$ can be obtained at the current density of ${10^8}~A/c{m^2}$.\\
\indent To demonstrate the applicability of above angular characterization for SOT, the SOT effective fields of the same device have been further characterized by another two well-established SOT quantification methods. The first one is the harmonic measurements with large sweeping magnetic field applied at ${\theta _H} = {86^ \circ }$ \cite{Qiu-SciRep}. Fig.~\ref{fig5}(a, b) show the best fits for the second harmonic Hall resistances in two schemes, which yield the values of ${H_L}=17.8~Oe$ and ${H_T}=13.9~Oe$ and they are very close to the results of angular characterization in Fig.~\ref{fig4}. Fig.~\ref{fig5}(c, d) show the second SOT characterization results using harmonic measurements at low $H$, for which $H$ is applied in the film plane \cite{Hayashi_PRB2014_Quantitativea}. By fitting the first and second harmonic Hall resistances, the values of ${H_L}=23.4~Oe$ and ${H_T}=21.5~Oe$ are obtained which are slightly larger than the above two methods. In Fig.~\ref{fig5}(e), we compare the results of these three methods that measured for the same device with the identical setup and AC current. While the applicability of angular characterization for SOT can be evidenced from the comparison plot, the larger SOT effective fields by small H fitting method can be also well explained by the thermoelectric contributions in the second harmonic signal, which is especially significant when the magnetization aligns around OOP direction as we will discuss in the following section of Cu/CoTb/Cu study.\\
\subsection{Cu/CoTb/Cu with Giant Thermal Effect
\protect
}
\begin{figure}
	\includegraphics[width=8.5cm]{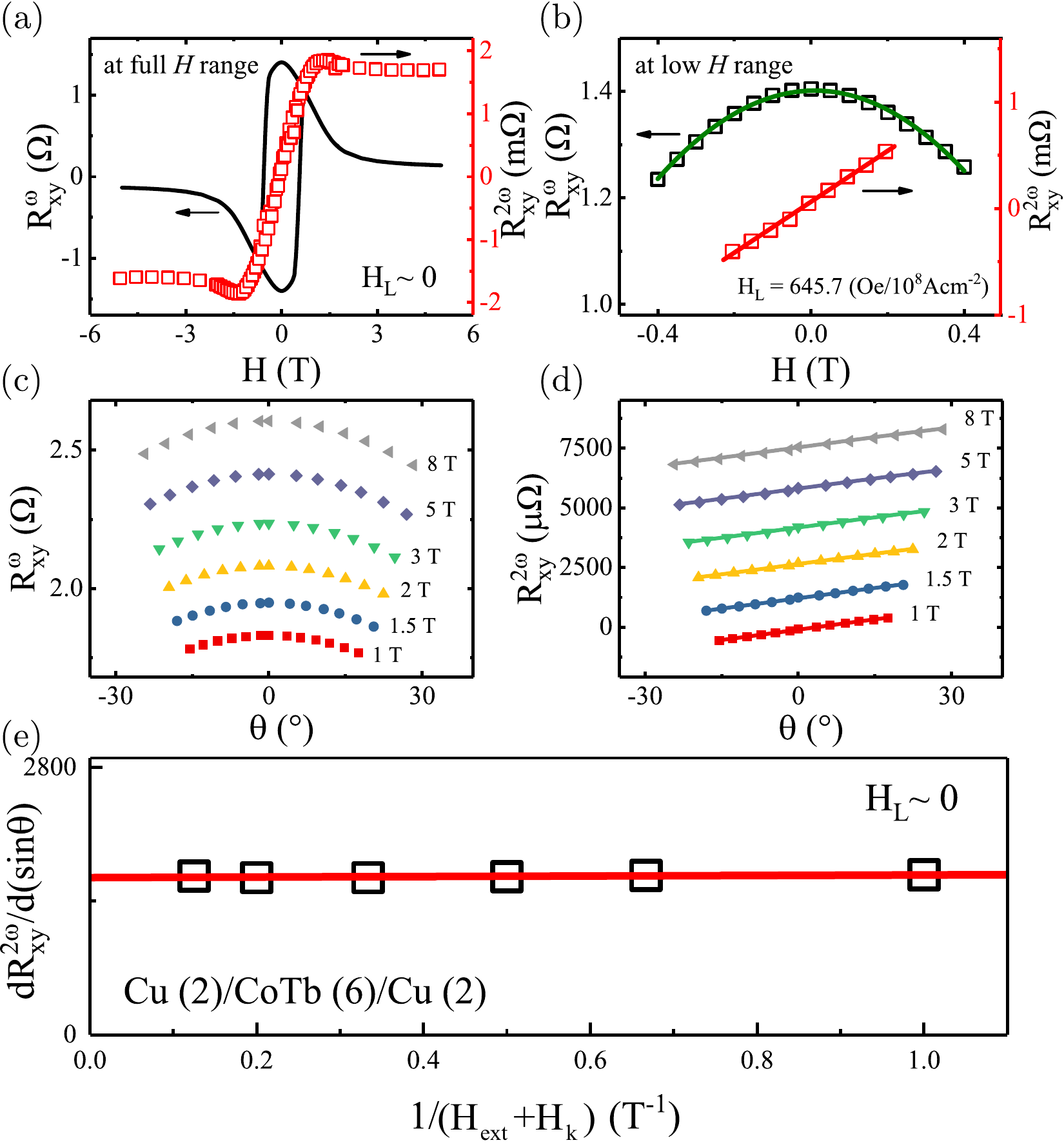}
	\caption{\label{fig6} (Color online) AC harmonic characterization of SOT for Cu/CoTb/Cu in the longitudinal scheme. Fitting the second harmonic signal at full $H$ range (a) and low $H$ range(b). (c)(d), Angular dependence of first and second harmonic Hall resistances with different constant external magnetic fields (${H_{ext}} \sim 1~T,~1.5~T,~2~T,~3~T,~5~T,~8~T$). The data are offset along the vertical axis for better clarification. (e) The SOT and thermoelectric contributions as a function of the inverse field ($1/({H_{ext}} + {H_k}$))}
\end{figure}
\indent Rare-earth metals of 4$f$-electron lanthanide series have been reported to accommodate significant spin Hall effect or SOT,
such as Tb has a large charge-to-spin efficiency of -0.18\cite{Wong_PRA2019,Reynolds_PRB2017}. However, some groups reported that the charge-to-spin efficiency of Rare-earth/3$d$-ferromagnet alloy or multilayer is negligibly small\cite{Han_PRL2017,Finley_PRA2016,Jiawei_NatM2019}. Here, we have characterized the SOT and thermoelectric effects in CoTb alloys by using the angular characterization to address these debates, as well as to reveal the artifact of thermoelectric effect in harmonic Hall measurements.\\
 \indent Fig.~\ref{fig6} show the measurement results for a Cu (2)/CoTb (6)/Cu (2) device, where Cu does not contribute the spin current and CoTb is ferrimagnetic alloy layer with perpendicular anisotropy. The magnetic field dependence of first and second harmonic Hall resistance with the sweeping magnetic field at the full and low $H$ range are shown in Fig.~\ref{fig6}(a, b). In the longitudinal scheme, we do not observe obvious peak or dip in the second harmonic loop at large $H$, which is the signature of SOT, in Fig.~\ref{fig6}(a). However, a steep slope of second harmonic Hall resistance is observed at small $H$ range, as shown in Fig.~\ref{fig6}(b). By fitting the first and second harmonic signals at small $H$ range, ${H_L} = 645.7~Oe/{10^8}Ac{m^{ - 2}}$ can be obtained, which is comparable or even larger than that in conventional Pt/FM films.\\
\indent We further characterize the device using the angular characterization method. Fig.~\ref{fig6}(c, d) show the angular dependence of the first and second harmonic Hall resistances measured by rotating the sample in the $xz$ plane around $z$-axis with a constant external magnetic field (${H_{ext}} \sim 1~T,~1.5~ T,~2~T,~3~T,~5~T,~8~T$). Fitting for the data with different ${H_{ext}}$ according to Eq.~(\ref{eq:dVdtheta}) are shown by solid curves in Fig.~\ref{fig6}(d). We noted that the slope $dR_{xy}^{2\omega }/d\theta$ is a constant that remains unchanged with ${H_{ext}}$, which can be more clearly seen in Fig.~\ref{fig6}(e). According to Eq.~(\ref{eq:dVdtheta}), the value of ${H_L}$ can be determined to be 0. Therefore, from the above studies, one can see that in Cu/CoTb/Cu sample the ANE contribution dominates the second harmonic signal at the small $H$ region, which possibly leads to fatal error in the SOT quantifications.\\
\subsection{CoNi/Pt with Weak PMA
\protect
}
\begin{figure}
	\includegraphics[width=8.5cm]{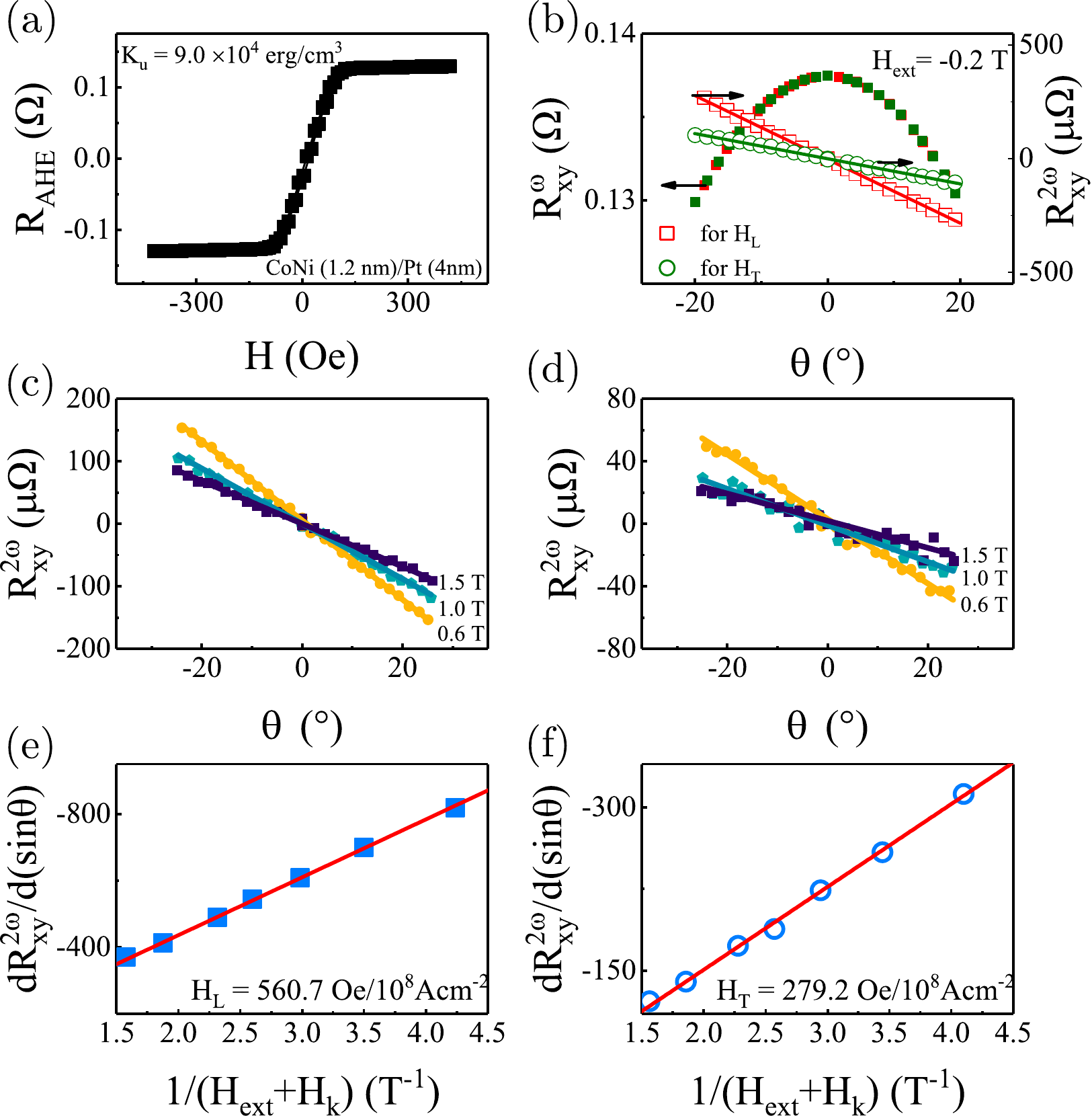}
	\caption{\label{fig7}(Color online) Measurements for a Ta/Cu/CoNi/Pt sample with weak PMA. (a) The OOP Hall hysteresis loop. (b) The angular characterization of the first and second harmonic Hall resistances with $H_{ext}=0.2~T$. (c)(d) Angular dependence of the second harmonic Hall resistance with different external magnetic fields ($H_{ext}\sim 0.6~T,~1.0~T,~1.5~T$) at the longitudinal and transverse schemes, respectively. The SOT and thermoelectric contributions as a function of the inverse magnetic field ($1/ ({H_{ext}} + {H_k})$) for $H_{L}$ (e) and $H_{T}$ (f).}
\end{figure}
\indent SOT research is still rapidly growing with great interests in exploring new materials with large change-to-spin efficiency, such as the topological insulator\cite{Mellnik-Nat2014,WangY-PRL2015,Kondou_NatP2016}, Weyl semimetal\cite{Ali_Nature2014,MacNeill_NatP2016}, and two-dimensional material\cite{Shuyuan_PRB2018,HePan_PRL2018}. However, it is not easy to form perpendicular magnetic heterostructure with these new materials for constructing a stable and efficient SOT device. Moreover, the PMA of sample can be unstable and degraded due to a harsh environment or frail sample structure. These place practical hurdles for SOT characterization. Thus, a versatile and comprehensive SOT characterization method that not limited by magnetic anisotropy is much needed to advance the SOT research.\\
\indent Here, we show that the angular characterization is capable for the SOT device with weak PMA. Fig.~\ref{fig7} shows the experiment results for the Ta (2)/Cu (3)/[Co (0.2)/Ni (0.4)]$_2$/Pt (4)/SiO$_2$ (3) (in nm). As shown in Fig.~\ref{fig7}(a), the sample has a weak PMA and uniaxial anisotropy energy ${K_{u}=9.0\times10^4~erg/cm^{3}}>0$. As a result, both the methods of harmonic measurement with large and small sweeping fields as in Fig.~\ref{fig5} \cite{Qiu-SciRep,Hayashi_PRB2014_Quantitativea} are not applicable here.\\
\indent We further apply the angular characterization for this sample. Fig.~\ref{fig7}(b) show the angular dependence of first and second harmonic Hall resistances with a constant external magnetic field (${H_{ext}} \sim 0.2~T$). The magnetization is in a saturation state while the field rotated around the z-axis. Fig.~\ref{fig7}(c, d) shows the second harmonic Hall resistance as a function of external magnetic field $H_{ext} \sim 0.6~T,~1.0~T,~1.5~T$. The inverse of magnetic field $1/(H_{ext}+H_{k})$ dependence of $dR_{xy}^{2\omega}/d(\sin\theta)$ are measured, as shown in Fig.~\ref{fig7}(e, f). Subsequently, the SOT effective field can be readily determined by fitting the data with Eq.~(\ref{eq:dVdtheta}). The longitudinal effective field $H_L = 560.7~Oe$ and the transverse effective field $H_T = 279.2~Oe$ are obtained at the current density of ${10^8}~A/c{m^2}$.\\
\section{\label{sec:level1}Conclusions}
\indent In conclusion, a new angular characterization method is established in this work to study SOT and thermoelectric effects in various magnetic heterostructures. The entanglement between $H_L$ and $H_T$, and the contributions from SOT and ANE are comprehensively considered. By accurately characterizing SOT and thermoelectric effects with high sensitivity, simple procedures and wide applicability, this work not only provides an important basis for SOT research and allows SOT explorations in more general material systems, but also bring great prospects for studying thermal spintronics.\\
\begin{acknowledgments}
This work was supported by the National Key R $\&$ D Program of China
Grand No. 2017YFA0303202 and 2017YFA0305300, the National Science Foundation of China Grant Nos. 11974260, 11674246, 51501131, 51671147, 11874283, 51801152, and 11774064, Natural Science Foundation of Shanghai Grant No. 17ZR1443700, 19ZR1478700, and the Fundamental Research Funds for the Central Universities.\\
\end{acknowledgments}
%
\appendix
\section{Principle and Formula Derivation}
\setcounter{equation}{0}
\renewcommand{\theequation}{A\arabic{equation}}
\subsection{\label{app:subsec} Current-Induced Second Harmonic Voltage}
\indent In Harmonic Hall measurements, we applied an AC current through the sample and the Hall voltage can be expressed as $V_{xy}=I_0R_{xy}(\theta+\Delta \theta (t),\varphi+\Delta \varphi (t))$, where, $\theta $ and $\varphi $ reflect the magnetization equilibrium direction determined by $\overrightarrow {{H_k}}$ and $\overrightarrow {{H_{ext}}}$, $\Delta \theta $ and $\Delta \varphi $ characterized the magnetization oscillation determined by $\overrightarrow {{H_{{I_0}}}} $ . So that the transverse resistance is a function about $\overrightarrow {{H_k}} $, $\overrightarrow {{H_{ext}}} $ and $\overrightarrow {{H_{{I_0}}}} $ can be rewritten as
\begin{equation}\label{eq:trigonometric}
\begin{split}
{R_{xy}} &= {R_{xy}}(\theta+\Delta \theta (t),\varphi+\Delta \varphi (t) )\\
&= {R_{xy}}(\overrightarrow {{H_k}}  , \overrightarrow {{H_{ext}}} , \overrightarrow {{H_{{I_0}}}(t)} )\\
\end{split}
\end{equation}\\
where, $\overrightarrow {{H_{{I_0}}}}  = \overrightarrow {{H_L}}  + \overrightarrow {{H_T}} $. For each equilibrium position with fixed $\theta$ and $\varphi$, the $\theta$ and $\varphi$ are fixed, ${R_{xy}}(t)$ is a function of $\overrightarrow {{H_{{I_0}}}}$ which can be further expanded to first-order approximation\cite{Avci-PRB2014}
\begin{equation}\label{eq:trigonometric}
\begin{split}
{R_{xy}}(t)\left| {_{\theta ,\varphi }} \right. \approx {R_{xy}}(\overrightarrow {{H_k}} ,\overrightarrow {{H_{ext}}} ) + {{d{R_{xy}}} \over {d\overrightarrow {{H_{{I_0}}}} }} \cdot \overrightarrow {{H_{{I_0}}}} \sin \omega t
\end{split}
\end{equation}
So the harmonic Hall voltage can be rewriten as 
\begin{equation}\label{eq:trigonometric}
\begin{split}
{V_{xy}}(t)  =& {I_0}\sin \omega t {R_{xy}(t)}\\
\approx &{I_0}{R_{xy}}(\overrightarrow {{H_k}}  , \overrightarrow {{H_{ext}}} )\sin \omega t + {I_0}{{d{R_{xy}}} \over {d\overrightarrow {{H_{{I_0}}}} }} \cdot \overrightarrow {{H_{{I_0}}}} {\sin ^2}\omega t\\
= &{I_0}[\frac{1}{2}\frac{{d{R_{xy}}}}{{d\overrightarrow {{H_{{I_0}}}} }} + {R_{xy}}(\overrightarrow {{H_k}}  , \overrightarrow {{H_{ext}}} )\sin \omega t \\
&- \frac{1}{2}\frac{{d{R_{xy}}}}{{d\overrightarrow {{H_{{I_0}}}} }} \cdot \overrightarrow {{H_{{I_0}}}} \cos 2\omega t]
\end{split}
\end{equation}
where ${V_{xy}^0}=\frac{{{I_0}}}{2}\frac{{d{R_{xy}}}}{{d\overrightarrow {{H_{{I_0}}}} }}\overrightarrow {{H_{{I_0}}}}$, ${V_{xy}^\omega}={I_0}{R_{xy}}(\overrightarrow {{H_k}}  , \overrightarrow {{H_{ext}}} )\sin \omega t$, and ${V_{xy}^{2\omega}}= - \frac{{{I_0}}}{2}\frac{{d{R_{xy}}}}{{d\overrightarrow {{H_{{I_0}}}} }}\overrightarrow {{H_{{I_0}}}} \cos 2\omega t$ are the zero, first, and the second harmonic components of the transverse voltage, respectively. With the inclusion of contribution from anomalous and planar Hall effects, the harmonic of transverse resistance can be expressed as
\begin{equation}\label{eq:trigonometric}
\begin{split}
R_{xy} &= {R_{AHE}}\cos (\theta ) + {R_{PHE}}{\sin ^2}(\theta )\sin (2\varphi)
\end{split}
\end{equation}
\indent By using the mathematical differential formula, for which $df(x,y) = {{\partial f} \over {\partial x}}dx + {{\partial f} \over {\partial y}}dy$, we have 
\begin{equation}\label{eq:trigonometric}
\begin{split}
d{{R}_{xy}} & =\frac{\partial {{R}_{xy}}}{\partial \cos \theta }d\cos \theta +\frac{\partial {{R}_{xy}}}{\partial \sin 2\varphi }d\sin 2\varphi  \\ 
& =\text{-}\sin \theta \frac{\partial {{R}_{xy}}}{\partial \cos \theta }d\theta +2\cos (2\varphi )\frac{\partial {{R}_{xy}}}{\partial \sin 2\varphi }d\varphi  \\ 
\end{split}
\end{equation}
\indent So the second harmonic of transverse resistance can be writen as
\begin{equation}\label{eq:R2w_D}
\begin{split}
 R_{xy}^{2\omega }  =& \frac{{V_{xy}^{2\omega}}}{I_0\cos 2 \omega t}\\
 =&-\frac{1}{2}\overrightarrow{{{H}_{{{I}_{0}}}}}\frac{d{{R}_{xy}}}{d\overrightarrow{{{H}_{{{I}_{0}}}}}}\\ 
 =&-\frac{1}{2}\overrightarrow{{{H}_{{{I}_{0}}}}}\frac{(-\sin \theta \frac{\partial {{R}_{xy}}}{\partial \cos \theta }d\theta +2\cos (2\varphi )\frac{\partial {{R}_{xy}}}{\partial \sin 2\varphi }d\varphi )}{d\overrightarrow{{{H}_{{{I}_{0}}}}}} \\ 
 =&\frac{1}{2}[{{R}_{AHE}}\sin \theta -{{R}_{PHE}}\sin 2\theta \sin (2\varphi )]\overrightarrow{H{}_{{{I}_{0}}}}\frac{d\theta }{d\overrightarrow{H{}_{{{I}_{0}}}}}\\
  &-{{R}_{PHE}}{{\sin }^{2}}\theta \cos 2\varphi \overrightarrow{H{}_{{{I}_{0}}}}\frac{d\varphi }{d\overrightarrow{{{H}_{{{I}_{0}}}}}} \\ 
\end{split}
\end{equation}
where, $\overrightarrow {{H_{{I_0}}}}$ can be decomposed into two components: the first one is ${H_{I_0}^\theta}$ that causing the magnetic moment to vibrate in the $\theta$ direction and $d{{H_{I_0}^\theta}}  = {(H_{ext}+H_k)}d\sin ({\theta} - {\theta'})$; another one is ${H_{I_0}^\varphi}$ that causing the magnetic moment to vibrate in the $\varphi$ direction and $d{{H_{I_0}^\varphi}}  = {H_{tot}}d\sin ({\varphi} - {\varphi'} )$. Obtained from the mathematical differential, we know that 
\begin{equation}\label{eq:math_D}
\begin{split}
\frac{d\theta }{dH_{{I_0}}^\theta } = \frac{1}{{(H_{ext}+H_k)}\cos (\theta  - \theta ')}\approx \frac{1}{H_{ext}+H_k} \\
\frac{d\varphi } {dH_{{I_0}}^\varphi } = \frac{1}{{(H_{ext}+H_k)}\cos (\varphi  - \varphi ')}\approx \frac{1}{H_{ext}+H_k}
\end{split}
\end{equation}
where the $\theta '$ and $\varphi '$ are the positions of magnetization driven by SOT effective fields. $\Delta \theta  = \theta  - \theta '$ and $\Delta \varphi  = \varphi  - \varphi '$ are the amplitudes of magnetization oscillation.\\
\indent We can simplify the second harmonic transverse resistance by combining Eq.~(\ref{eq:R2w_D}) and~(\ref{eq:math_D}) for the two measurement schemes.\\
\indent (1)	For the longitudinal scheme: the external magnetic field is applied in the $xz$ plane, $\varphi  \approx 0^\circ $, $H_{{I_0}}^\theta  = {H_L}$ and $H_{{I_0}}^\varphi  = {H_T}$.\\
So we have
\begin{equation}\label{eq:assu_long}
\begin{split}
{\sin \varphi  \approx 0};{\sin {2\varphi}  \approx 0};{\cos \varphi  \approx 1};{\cos {2\varphi}  \approx 1}
\end{split}
\end{equation}
The second harmonic term can be simplified to
\begin{equation}\label{eq:SOT_L}
\begin{split}
R{_{xy}^{2\omega }}=  &- \frac{1}{2}[{R_{AHE}}\frac{{d\cos \theta }}{{d{\theta}}} \cdot \frac{{H_{{I_0}}^\theta }}{{{H_{ext}+H_k}}} \\
&+ {R_{PHE}}{\sin ^2}\theta \frac{{d\sin (2\varphi )}}{{d{\varphi}}} \cdot \frac{{H_{{I_0}}^\varphi }}{{{H_{ext}+H_k}}}]\\
=  &\frac{1}{2}{R_{AHE}}\sin \theta \frac{{{H_L}}}{{{H_{ext}+H_k}}}- {R_{PHE}}{\sin ^2}\theta \frac{{{H_{T}}}}{{{H_{ext}+H_k}}}\\
= &\frac{R_{AHE}}{2(H_{ext}+H_k)}(H_L-\frac{R_{PHE}}{R_{AHE}}H_T \sin \theta)\sin \theta\\
\approx &\frac{1}{2}{R_{AHE}}\frac{H_L}{H_{ext}+H_k}\sin \theta
\end{split}
\end{equation}
\indent (2)	For transverse scheme: the external magnetic field is applied in the $yz$ plane, $\varphi  \approx 90^\circ $, $H_{{I_0}}^\theta  = {H_{T}}$ and $H_{{I_0}}^\varphi  = {H_L}$\\
So we have
\begin{equation}\label{eq:assu_trans}
\begin{split}
{\sin \varphi  \approx 1};{\sin {2\varphi}  \approx 0};{\cos \varphi  \approx 0};{\cos {2\varphi}  \approx -1}
\end{split}
\end{equation}
The second harmonic term can be rewritten as
\begin{equation}\label{eq:SOT_T}
\begin{split}
R{_ {xy} ^{2\omega }} =  &- \frac{1}{2}[{R_{AHE}}\frac{{d\cos \theta }}{{d{\theta }}} \cdot \frac{{H_{{I_0}}^\theta }}{{{H_{ext}+H_k}}} \\
&+ {R_{PHE}}{\sin ^2}\theta \frac{{d\sin (2\varphi )}}{{d{\varphi}}} \cdot \frac{{H_{{I_0}}^\varphi }}{{{H_{ext}+H_k}}}]\\
= & \frac{1}{2}{R_{AHE}}\sin \theta \frac{{H_{T}}}{{{H_{ext}+H_k}}} + {R_{PHE}}{\sin ^2}\theta \frac{{{H_L} }}{{{H_{ext}+H_k}}}\\
= &\frac{R_{AHE}}{2(H_{ext}+H_k)}(H_T+\frac{R_{PHE}}{R_{AHE}}H_L \sin \theta)\sin \theta\\
\approx &\frac{1}{2}{R_{AHE}}\frac{H_T}{H_{ext}+H_k}\sin \theta
\end{split}
\end{equation}
\indent (3) Fitting Proximation and PHE correction: Since $\sin^2 \theta\approx 0$ and $\xi=\frac{R_{PHE}}{R_AHE}$ is negligible, we ignore the influence of PHE on second harmonic signals, for which: $H_L-\frac{R_{PHE}}{R_{AHE}}H_T \sin \theta\approx H_L$ and $H_T+\frac{R_{PHE}}{R_{AHE}}H_L \sin \theta\approx H_T$. However, for the more strict and accurate analysis by including the entanglement between $H_L$ and $H_T$ through PHE, the correction equation can be express as:
\begin{equation}\label{eq:PHE_recorrect2}
\begin{split}
{H_L^{'} = H_L + 2\xi{H_T}\sin \theta }\\
{H_T^{'} = H_T - 2\xi{H_L}\sin \theta }.
\end{split}
\end{equation}
When $\theta=0$, $H_L^{'}=H_L$ and $H_T^{'}=H_T$. For $\theta\ne0$, we can correct the SOT effective field with the influence of PHE at each $\theta$.\\
\subsection{\label{app:subsec} Thermoelectric Contribution}
\indent In harmonic measurements, an in-plane current injection induces an in-plane and an out of plane thermal gradient, which contribute significant signal in the second harmonic term. In this part, we will show the model of thermoelectric contributions. Fig.~\ref{fig3} (a) shows the finite element analysis of the current induced thermal gradient, where the structure of the film is Si/${\rm{SiO}}_{\rm{2}}$/HM/FM/${\rm{SiO}}_{\rm{2}}$. The simulation parameter: 1, the isotropic thermal conductivity ($\sigma$) of each layer, ${\sigma _{Si}} = {\sigma _{HM}} = \sigma {}_{FM} = 60$~($Wm^{-1}C^{\circ-1}$), ${\sigma_{SiO_{2}}} = 10$~($Wm^{-1}C^{\circ-1}$); 2, Film coefficient= $10~(Wm^{-1}C^{\circ-1})$; 3, Ambient temperature, $22 C^\circ$; 4, The temperature load: $T_{FM} = 50~C^\circ$, $T_{HM}=75~C^\circ$, $T_{Si/SiO_2}=23.5~C^\circ$; 5, Mesh, default; There are two directions of thermal gradient induced by the injected current. First, we define the vector of the thermal gradient are ${\overrightarrow {\nabla T} _{ip}} = \nabla {T_{ip}}(1,0,0)$ (the in-plane part), ${\overrightarrow {\nabla T} _{oop}} = \nabla {T_{oop}}(0,0,1)$ (the out of plane part), respectively. Moreover, we have the magnetization direction is $\overrightarrow m  = (\sin \theta \cos \varphi ,\sin \theta \sin \varphi ,\cos \theta )$. When the $\overrightarrow m$ experiences a thermal gradient $\overrightarrow {\nabla T}$, the ANE voltage is produced and its direction obeys Eq.~(\ref{eq:thermal ANE}). Thus, the second harmonic ANE contribution term can be established as follows.\\
\indent a.	The in-plane ANE contribution for second harmonic term
\begin{equation}\label{eq:trigonometric}
\begin{split}
R_{oop,ANE}^{2\omega } = &{I_0}\alpha \overrightarrow {\nabla {T_{oop}}}  \times \overrightarrow m  \cdot \overrightarrow y \\
= &{I_0}\alpha \nabla {T_{oop}}(0,0,1) \\
&\times (\sin \theta \cos \varphi ,\sin \theta \sin \varphi ,\cos \theta ) \cdot (0,1,0)\\
= &{I_0}\alpha \nabla {T_{oop}}\sin \theta \cos \varphi .
\end{split}
\end{equation}\\
\indent b.	The out of plane ANE contribution for second harmonic term
\begin{equation}\label{eq:trigonometric}
\begin{split}
R_{ip,ANE}^{2\omega } = &{I_0}\alpha \overrightarrow {\nabla {T_{ip}}}  \times \overrightarrow m  \cdot \overrightarrow y \\
= &{I_0}\alpha \nabla {T_{ip}}(1,0,0) \\
&\times (\sin \theta \cos \varphi ,\sin \theta \sin \varphi ,\cos \theta ) \cdot (0,1,0)\\
=  &- {I_0}\alpha \nabla {T_{ip}}\cos \theta .
\end{split}
\end{equation}
As shown in Fig.~\ref{fig3}(b, d), in the longitudinal scheme, we have
\begin{equation}\label{eq:thermal_L}
\begin{split}
R_{\nabla T}^{2\omega } = {I_0}\alpha (\nabla {T_{oop}}\sin \theta  - \nabla {T_{ip}}\cos \theta )
\end{split}
\end{equation}
As shown in Fig.~\ref{fig3}(c, e), in the transverse scheme, we have
\begin{equation}\label{eq:thermal_T}
\begin{split}
R_{\nabla T}^{2\omega } =  - {I_0}\alpha \nabla {T_{ip}}\cos \theta
\end{split}
\end{equation}
\indent Combining Eq.~(\ref{eq:SOT_L}),~(\ref{eq:SOT_T}),~(\ref{eq:thermal_L}) and~(\ref{eq:thermal_T}), with the inclusion of SOT effects and thermoelectric contribution, the second harmonic term $R_{xy}$ can be expressed as
For the longitudinal scheme:
\begin{equation}\label{eq:trigonometric}
\begin{split}
R_{xy }^{2\omega } =  & \frac{1}{2}({R_{AHE}}{\frac{{{H_L}}}{{{H_{ext}+H_k}}}} + 2{I_0}\alpha \nabla {T_{oop}})\sin \theta \\
& - {I_0}\alpha \nabla {T_{ip}}\cos \theta
\end{split}
\end{equation}
For the transverse scheme:
\begin{equation}\label{eq:trigonometric}
\begin{split}
R_{ xy }^{2\omega } = &\frac{1}{2}{R_{AHE}}\frac{{{H_{T}}}}{{{H_{ext}+H_k}}}\sin \theta- {I_0}\alpha \nabla {T_{ip}}\cos \theta
\end{split}
\end{equation}
where the influence of PHE can be recorrected by the Eq.~(\ref{eq:PHE_recorrect2}).
%


\end{document}